\documentclass[aps,prd,showpacs,notitlepage,nofootinbib,superscriptaddress,floatfix,showkeys,twocolumn]{revtex4-1}
\pdfoutput=1
\usepackage{graphicx}
\usepackage{amsmath,calligra,mathrsfs}
\usepackage{amsfonts}
\usepackage{amssymb}
\usepackage{bm}
\usepackage{physics}
\usepackage{tabularx}
\usepackage{multirow}
\usepackage{longtable}
\usepackage{array}
\usepackage{float}
\usepackage{xcolor, soul}
\usepackage{epstopdf}
\usepackage{graphicx}
\usepackage{amsmath}
\usepackage{amsfonts}
\usepackage{amssymb}
\usepackage{xcolor, soul}
\usepackage{epstopdf}
\usepackage{float}
\usepackage{subfigure}
\usepackage{float}
\usepackage{subfigure}
\usepackage{hyperref}
\usepackage{ulem}
\hypersetup{
     colorlinks   = true,
     citecolor    = red,
     linkcolor    = blue,
     urlcolor     = blue,
}\DeclareGraphicsExtensions{.eps} 
\def\beq{\begin{equation}}
\def\eeq{\end{equation}}
\def\bea{\begin{eqnarray}}
\def\eea{\end{eqnarray}}
\def\be{\begin{equation}}
\def\ee{\end{equation}}

\def\bse{\begin{subequations}}
\def\ese{\end{subequations}}

\def\aend{a_{\rm end}}

\def\nuth{\nu_{\rm R}^3}

\def\are{a_{\rm re}}
\def\aend{a_{\rm end}}
\def\tre{T_{\rm re}}
\def\kre{k_{\rm re}}
\def\kend{k_{\rm end}}

\def\mp{M_{\rm p}}
\def\wphi{w_{\rm \phi}}
\def\hend{H_{\rm end}}

\graphicspath{{./figs/}}
\begin{document}

\title{Gravitational neutrino reheating}

\author{Md Riajul Haque}%
\email{riaj.0009@gmail.com}
\affiliation{Centre for Strings, Gravitation and Cosmology,
Department of Physics, Indian Institute of Technology Madras, 
Chennai~600036, India}
\author{Debaprasad Maity}
\email{debu@iitg.ac.in}
\author{Rajesh Mondal}%
\email{mrajesh@iitg.ac.in}
\affiliation{%
	Department of Physics, Indian Institute of Technology Guwahati, Guwahati 781039, Assam, India}%

\date{\today}

\begin{abstract}
Despite having important cosmological implications, the reheating phase is believed to play a crucial role in both cosmology and particle physics model building. Conventional reheating models primarily relies on arbitrary coupling between inflaton and massless fields which lacks robust prediction. In this submission, we propose a novel reheating mechanism where the particle physics model, namely the Type-I seesaw is shown to play a major role in the entire reheating process, and inflaton is assumed to be free from arbitrary coupling. To the best of our knowledge, it is the first reheating model of the kind which, besides being successful in resolving the well-known neutrino mass and baryon asymmetry problems, constraints large class of inflation models, offers successful reheating, predicts distinct primordial gravitational wave spectrum and non-vanishing lowest active neutrino mass. Our novel mechanism opens up a new avenue of integrating particle physics and cosmology in the context of reheating. 
\end{abstract}

\maketitle
\section{Introduction}
 If the dominant decay channel of the inflaton is mediated by gravitons, can the universe be populated by radiation? Such a question has gained significant interest \cite{Haque:2022kez,Clery:2021bwz,Clery:2022wib,Dimopoulos:2018wfg,Nakama:2018gll}, particularly in the context of the model-independent reheating scenario. Recently proposed gravitational reheating (GRe) \cite{Haque:2022kez,Haque:2023yra} is one such interesting example, where the inflaton equation of state $(w_{\rm\phi})$ and inflationary energy scale are the only controlling parameter. Due to Planck suppression the gravitational decay of inflaton is expected to be maximal near the end of the inflation due to the large inflaton amplitude. Consequently the produced radiation redshifts as $\rho_R \sim a^{-4}$ from the very beginning of reheating while inflaton goes as $\rho_{\phi} \sim a^{-3(1+\wphi)}$. This essentially renders the inflaton equation of state ($\wphi$) to be stiff $w_\phi>0.65$ to achieve reheating temperature above BBN temperature $T_{\rm BBN}\sim 4$ MeV \cite{Kawasaki:2000en,Sarkar:1995dd,Hannestad:2004px}. Further, $w_\phi>0.65$ condition predicts blue tilted Primordial Gravitational Waves (PGWs) spectrum, and that is severely constrained by total massless degrees of freedom at BBN \cite{Mishra:2021wkm,Haque:2021dha,Pagano:2015hma,Yeh:2022heq,Planck:2018vyg}. This fact turned the GRe scenario viable only for a kination-like equation of state $\wphi\sim 1$ \cite{Haque:2022kez}.
 
It is in this parlance we invoke a 
particle physics-inspired Type-I seesaw extension of the standard model (SM) and investigate its impact on the aforesaid GRe scenario. Type-I seesaw \cite{Gell-Mann:1979vob,see1,see2,JWF1,JWF2,Datta:2021elq} is a minimal extension in the SM neutrino sector, which is known to generate active neutrino mass \cite{Super-Kamiokande:1998kpq,SNO:2002tuh,K2K:2002icj} and baryon asymmetric universe \cite{Planck:2018vyg}. In this submission, we show a new attributes to this model that it can also successfully reheat our universe without invoking any new physics in the inflaton sector. We call it gravitational neutrino reheating\footnote{Our present proposal has great similarity with the recently proposed primordial black hole (PBH) reheating \cite{RiajulHaque:2023cqe,Haque:2023lzl} and here instead of BH we have massive right-handed neutrino as an extra component.}. Additionally, unlike the GRe scenario, gravitational neutrino reheating is shown to be consistent with the BBN bounds of PGWs and further predicts the non-vanishing mass of the lowest active neutrino.

The paper is organized as follows: In Sec.\ref{sc2}, we briefly discuss our 
 reheating framework and its connection with the Type-I seesaw model. In Sec.\ref{sc3} and \ref{sc4}, we review the gravitational production of RHNs and purely gravitational reheating (GRe) scenarios, respectively. In Sec.\ref{sc5}, we discuss in detail the gravitational neutrino reheating. In Sec.\ref{sc6}, we analyze the possible constraints of leptogenesis. Finally, conclude in Sec.\ref{scl}
\section{Type-I seesaw and reheating framework}{\label{sc2}}
Type-I seesaw  consists of three right-handed SM singlet massive Majorana neutrinos $\nu_R^i(i=1,2,3)$ with the following lepton number violating Lagrangian 
\begin{equation}
    \mathcal{L}= -\frac{1}{2}\,M_{\rm i}\,\overline{\nu^{\rm c i}_{\rm R}}\,\nu^{\rm i}_{\rm R}-y_{ij}\bar L_{\rm i}\Tilde{H}{\nu}^{\rm j}_{\rm R}+h.c.,
\end{equation}
where $\tilde {H} = i\sigma_2 H^*$, with $H (L)$ is the SM Higgs (lepton) doublet. Except usual inflaton parameters, Yukawa coupling matrix $y_{ij}$, and Majorana mass $M=\mbox{diag}(M_1, M_2, M_3)$ are the only beyond the SM parameters. After the inflation, all the particles are produced from the inflaton condensate through gravitational interaction only. For the reheating dynamics to occur, therefore, the relevant Boltzmann equations are,
\begin{subequations}
\begin{align}
& \dot{\rho_{\rm\phi}}+3H(1+w_\phi)\rho_{\rm\phi}+\Gamma^{T}_{\phi}(1+w_\phi)\rho_{\rm\phi}=0 ,\label{rhophi}\\
&\dot{\rho}_{\rm r}+4H\rho_{\rm r}-\Gamma_{\rm\phi\phi\rightarrow hh}(1+w_{\rm\phi})\rho_{\rm\phi}-\Gamma_{i}\, M_i\, n_{i}=0\,,\label{rhor} \\
&\dot{n}_{i}+3Hn_{i}-R^{i}_{\rm\phi}+\Gamma_{\rm i}\,n_{i}=0\,, \label{rhoN3}
\end{align}
\end{subequations}
where $\rho_{\rm \phi}$, $\rho_{\rm r}$ are the energy density of the inflaton, radiation respectively. $n_{i}$'s are the number density of the RHNs. $R^{i}_{\phi}$'s are the gravitational production rate of the RHNs \cite{Clery:2021bwz} . $\Gamma^{\rm T}_{\rm\phi}$ is the total inflaton decay width, $\Gamma_{\rm\phi\phi\rightarrow hh}$ is the gravitational decay width \cite{Ahmed:2022tfm} of inflaton to SM Higgs, and $\Gamma_{i}={(y^\dagger y)_{\rm ii}}M_{i}/(8\pi)$'s are the decay width of the RHNs. Out of all the parameters, the following three
$\{\beta = \sqrt{(y^\dagger y)_{\rm 33}}, M_3, w_\phi\}$, 
related to $\nu^3_R$ and inflaton will be shown to control the reheating dynamics. The remaining parameters will give rise to active neutrino mass and baryogenesis.


Type-I seesaw model consists of three copies of right-handed neutrinos (RHNs); among them two RHNs (namely $\nu^1_{\rm R},\nu^2_{\rm R}$) will be shown to explain both the light neutrino mass and baryon asymmetry of the universe  \cite{asaka,Campbell:1992hd,Kaneta:2019yjn,Bernal:2021kaj,Datta:2023pav,Hahn-Woernle:2008tsk}. And the third one $(\nu^3_R)$ is assumed to be long-lived with the decay width, $\Gamma_{3} \simeq \beta^2 M_3/(8\pi) \ll \Gamma_1,\Gamma_2$, where $\beta^2\simeq(m_1M_3/v^2)$. It will play the main role for successful reheating. Since $\beta^2 \propto m_1$, the non-vanishing $\Gamma_3$ immediately suggests that 
neutrino-controlling reheating indeed requires the lowest active neutrino mass to be non-vanishing. Here, we are using the normal hierarchy of the active neutrino masses $(m_1 < m_2<m_3)$. The SM Higgs vacuum expectation value $v=174$ GeV, and $m_3=0.05$ eV.

\section{RHN$\mbox{s}$ production from inflaton}{\label{sc3}}
Before its decay, the evolution of RHN number density ($n_{\rm i}$) is expected to be governed by the Boltzmann equation,
\be
\dot{n}_{\rm i}+3Hn_{\rm i}=R^{\rm i}_{\rm\phi} = \frac{ \rho^2_{\rm \phi}}{4\pi\,\gamma^2 M^4_{\rm p}}\frac{M^2_{i}}{m^2_{\rm\phi}}\Sigma_i\,, \label{numN}
\ee
where, $R^{\rm i}_{\rm\phi}$ the production rate for $\nu^{\rm i}_{\rm R}$ from inflaton scattering mediated by gravity \cite{Clery:2021bwz}. The factor $\Sigma_i\sim 10^{-2}$, is nearly constant and $\gamma$ is related with the oscillating inflaton condensate\footnote{ $\gamma$ is the frequency of the oscillation in the unit of $m_\phi$, $\gamma=\sqrt{\frac{\pi\, n}{2n-1}}\frac{\Gamma\left(\frac{1}{2}+\frac{1}{2n}\right)}{\Gamma\left(\frac{1}{2n}\right)}$.} at the minimum of its potential. For inflation, we consider the $\alpha$-attractor $E$-model potential \cite{Kallosh:2013hoa,Ferrara:2014cca,Ueno:2016dim},
$
V(\phi)=\Lambda^4\left(1-e^{-\sqrt{2/(3\,\alpha)}\,\phi/\mp}\right)^{2n}$, 
where $\Lambda \sim8\times 10^{15}$ GeV is the inflation mass-scale fixed by the CMB power spectrum . The parameters $(\alpha,n) $ control the shape of the potential. The inflaton equation of state $w_\phi$ is directly related with $n$, via the relation $\wphi=(n-1)/{(n-1)}$ \cite{Garcia:2020eof}.
For this model, the time-dependent inflaton mass can be expressed as 
\cite{Garcia:2020eof,btm1}
$
m_\phi^2 \sim {(m_\phi^{\rm end})}^2 ({a}/{a_{\rm end}})^{-6\,w_\phi}$, 
 $m_\phi^{\rm end}$ is the inflaton mass defined at the end of the inflation,
\be\label{massend}
m_\phi^{\rm end}\simeq  \frac{\sqrt{4n\,\left(2n-1\right)}}{\sqrt{3\,\alpha}}\frac{\Lambda^{\frac{2}{n}}}{M_{\rm p}}\left(\rho_\phi^{\rm end}\right)^{\frac{n-1}{2n}}\,.
\ee
We consider $\alpha=1$ throughout unless otherwise stated. During the initial period of reheating inflaton is typically the dominant energy component and neglecting the decay term in Eq.(\ref{rhophi}), $\rho_{\rm\phi}$ evolves as, 
$\rho_\phi(a)= 3M_{\rm p}^2H_{\rm end}^2(a/{a_{\rm end}})^{-3(1+w_\phi)}$,  
 $\hend$ being the Hubble constant at the end of the inflation. $\mp=2.4\times10^{18}$ GeV is the reduced Planck
mass. Solving Eq.\ref{numN} one obtains, 
\begin{equation}\label{Nis}
\begin{aligned}
    n_{\rm i}(a)\simeq \frac{3H^3_{\rm end}M^2_{\rm i}\Sigma_{\rm i}}{2\pi(1-w_\phi)(\gamma\, m^{\rm end}_\phi)^2} \left(\frac{a}{\aend}\right)^{-3},
    \end{aligned}
\end{equation}
where the effective initial number density of RHNs at the end of inflation is 
\begin{equation}
\begin{aligned}
    n_i^{\rm end}
    \simeq 10^{35}\frac{1}{2\gamma^2}\left(\frac{H_{\rm end}}{10^{13}}\right)^3\times \left(\frac{M_i}{10^{12}}\right)^2\left(\frac{10^{13}}{m^{\rm end}_\phi}\right)^2\,.
    \end{aligned}
\end{equation}
In the above expression, all are written in $\mbox{GeV}$ and assume $w_\phi =0$.
 \begin{figure}[t]
\centering
\includegraphics[width=\columnwidth]{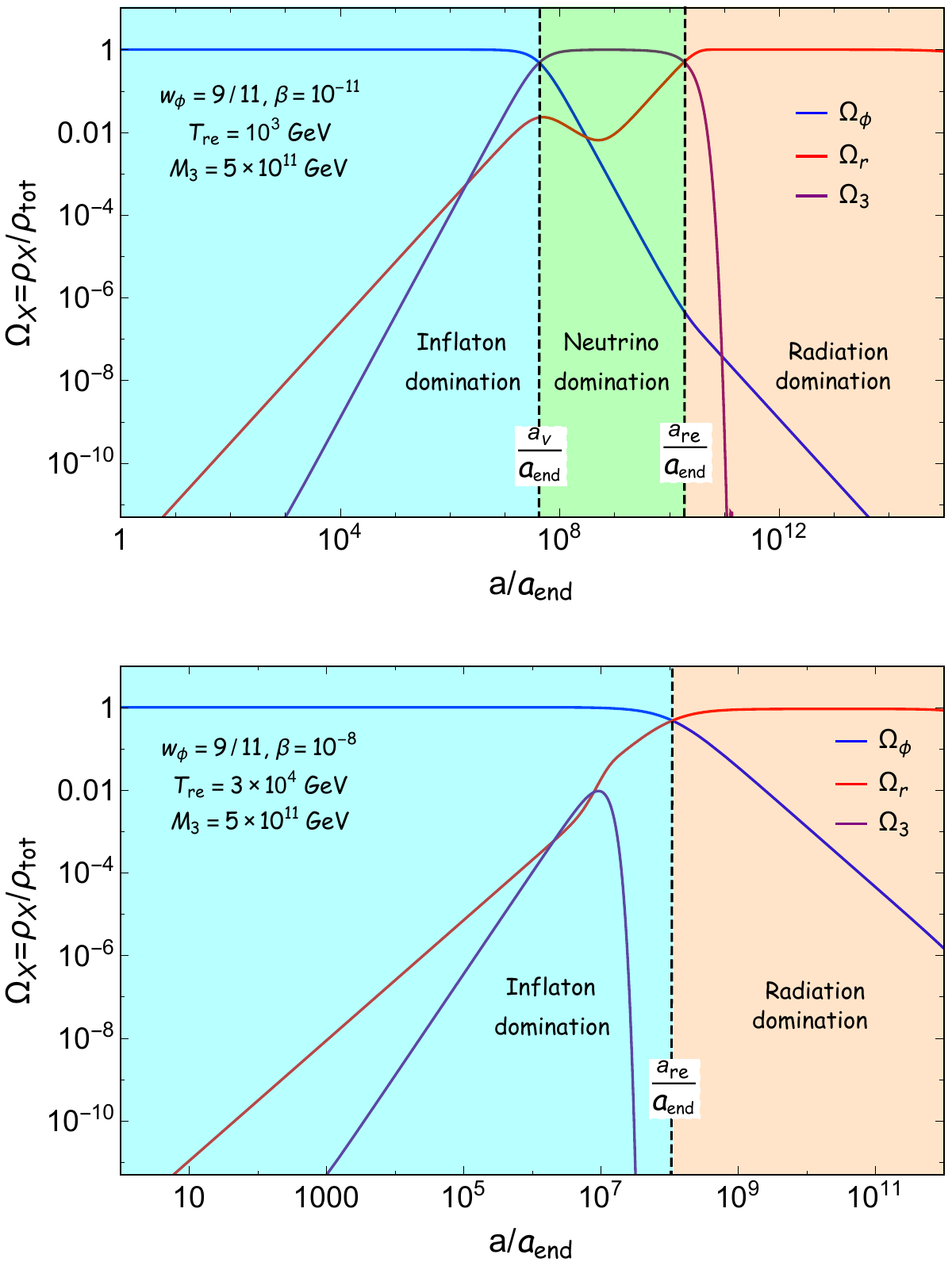}
\caption{The evolution of the normalized energy densities $\Omega_{\rm X}=\rho_{\rm X}/\rho_{\rm tot}$ as a function of the scale factor for $w_\phi=9/11$
with $M_3=5\times10^{11}$ GeV. The top plot corresponds to the
neutrino dominating case ($\beta<\beta^{\rm c}_{\rm\nu}$), and the bottom plot corresponds to the neutrino heating case ($\beta>\beta^{\rm c}_{\rm\nu}$)}
          \label{density}
      \end{figure}
With this ingredient in hand, we now discuss various causes and conditions of different phases of reheating.

\section{ Gravitational Reheating (${\rm GRe}$)}{\label{sc4}}
 In this scenario, the radiation dynamics is assumed to be solely controlled by inflaton via gravitational scattering, and using Eq.\ref{rhor}, one obtains,
\begin{equation}\label{rhorg}
    \rho^\phi_{\rm r}(a)\simeq\frac{9\gamma H^3_{\rm end}m^{\rm end}_{\rm\phi}}{10(1+15w_{\rm\phi})\pi}\left(\frac{a}{a_{\rm end}}\right)^{-4} .
\end{equation}
For the exact expression see, Ref.\cite{Clery:2021bwz}.
The radiation produced from the neutrino decay can be solved by combining Eq.\ref{Nis} and Eq.\ref{rhor} as 
\begin{equation}\label{rhorphi}
    \rho^\nu_{\rm r}(a)\simeq\frac{2\beta^2_{\rm 3}M^2_{\rm 3}\,n^{\rm end}_{\rm3}}{8\pi(5+3w_{\rm\phi})H_{\rm end}}\left(\frac{a}{a_{\rm end}}\right)^{-\frac{3(1-w_{\rm\phi})}{2}} .
\end{equation}
Using this we, therefore, define a critical decay width, 
\begin{equation}
\begin{aligned}
    \beta^{c}_\phi\simeq&\left(\frac{\gamma}{3\pi(1+15w_{\rm\phi}) }\frac{m^{\rm end}_{\rm\phi}}{M_{\rm p}}\right)^{\frac{-3(1+w_\phi)}{4}}\left(\frac{M_{\rm p}}{H_{\rm end}}\right)^{\frac{7+9w_\phi}{4}}\\
  &\quad\times\left(\frac{n_{\rm 3}^{\rm end}}{M^3_{\rm p}}\right)^{\frac{3(1+w_{\rm\phi})}{4}}\left(\frac{M_{\rm p}}{M_3}\right)^{\frac{-(1+3w_{\rm\phi})}{4}}.
    \end{aligned} .
\end{equation}
Particularly, if the coupling is very large satisfying $\beta >\beta^c_\phi$, immediately after its production, neutrinos are expected to decay and cannot lead the reheating process. Therefore, GRe would prevail. However, we mainly discuss the alternative possibilities.

\section{Gravitational Neutrino reheating}{\label{sc5}}
As mentioned earlier $\Gamma_3 \ll \Gamma_1,\Gamma_2$, and hence $\nu^3_{\rm R}$ is a long-lived particle compared to the other two. Because of that $\nu^3_{\rm R}$ can compete with inflaton and alter the reheating process. Hence we have three major players to deal with; inflaton and $\nu^3_{\rm R}$ and radiation. The energy density of the $\nu^3_{\rm R}$ before its decay evolves as, 
\begin{equation}\label{rhoN3s}
\begin{aligned}
    \rho_{3}(a)&\simeq\frac{3H^3_{\rm end}\,M^3_{3}}{2\pi(1-w_\phi)(10\,\gamma\,m^{\rm end}_\phi)^2} \left(\frac{a}{\aend}\right)^{-3}.
    \end{aligned}
\end{equation}
The immediate observation from Eq.(\ref{rhoN3s}) is that for $w_{\rm\phi}>0$, since inflaton dilutes faster than the $\nuth$, it starts to dominate at $a=a_{\nu}$ where $\rho_{\rm\phi}\sim\rho_{3}$.
\begin{equation}\label{AphiN}
\begin{aligned}
    &\frac{a_\nu}{\aend} \simeq\left(200\,\pi\,(1-w_{\rm\phi})\,\gamma^2\right)^{1/3w_\phi}\left(\frac{M_{\rm p}}{H_{\rm end}}\right)^{1/3w_\phi}\\
    &\quad\quad\quad\quad\times\left(\frac{m^{\rm end}_\phi}{M_{\rm p}}\right)^{2/3w_\phi}\left(\frac{M_{\rm p}}{M_3}\right)^{1/w_\phi}.
    \end{aligned}
\end{equation}
 Once $\nu^3_{\rm R}$ dominates over the inflaton, its subsequent decay into SM fields would populate the universe. Since it dominates at $a_\nu$, $\nu_R^3$ decay before this point should be subdominant. Hence, one indeed expects another critical $\beta^c_\nu$ value below which only $\nu^3_{\rm R}$ can dominate over the inflaton, and it must be at the instant when $\Gamma_{\rm 3} = \beta^2 M_{3}/(8\pi) \sim H(a_{\nu})$. This condition immediately gives us,
\begin{equation}
\begin{aligned}
\beta^{\rm c}_\nu 
\simeq 22.3\left(\frac{H_{\rm end}}{10^{13}}\right)^{1/2}\left(\frac{5\times10^{11}}{M_3}\right)^{1/2}
\left(\frac{a_\nu}{a_{\rm end}}\right)^{-\frac{3(1+w_\phi)}{4}}
\end{aligned}
\end{equation}
For example, we get $\beta^{\rm c}_\nu \simeq2\times 10^{-9}$ for $w_\phi=9/11$ and $M_{\rm 3}=5\times 10^{11}$ GeV. In the bottom plot of Fig.(\ref{density}), we indeed see that $\nu^3_{\rm R}$ domination ceases to exist when $\beta = 10^{-8}>\beta^{\rm c}_\nu$. Using these parameters, we can say from Eq.(\ref{AphiN}) that our universe becomes neutrino-dominated when $a_{\rm\nu}/a_{\rm end}\simeq 4\times 10^{7}$, roughly after 17 e-folding number from the beginning of reheating. We, therefore, have two distinct possibilities for the reheating, 
\subsection{Neutrino dominating case:\,$\beta \leq \beta^{\rm c}_\nu$}
As stated earlier, for this condition $\nu^3_{\rm R}$ dominates over the inflaton after $a=a_{\rm\nu}$, and universe becomes matter dominated as $\rho_3 \propto a^{-3}$ (see Eq.\ref{rhoN3s}). Hence, reheating will end once $\nu_{\rm R}^3$ decays completely into radiation at $\Gamma_3\sim H$. During this period, the Hubble parameter evolves as 
$H(a)=H_{\rm end}(a_{\rm\nu}/a_{\rm end})^{-3(1+w_{\rm\phi})/2} ({a}/{a_\nu})^{-{3}/{2}}$ for $a \geq a_{\rm\nu}$, 
Utilizing this in Eq.(\ref{rhor}), radiation evolves as  
  \begin{equation}
    \rho_{\rm r}(a \geq a_{\rm\nu})\simeq\frac{\beta^2M^2_{3}\,n_{\rm 3}(a_{\rm\nu})}{20\,\pi\,H(a_{\rm\nu})} \left(\frac{a}{a_{\rm\nu}}\right)^{-3/2} .
\end{equation}
The evolution of radiation energy density is independent of $w_{\rm\phi}$ as expected. However, For $(a<a_{\rm\nu})$ radiation is populated by the inflaton via gravitational decay.
\begin{figure}[t]
\centering
\includegraphics[width=\columnwidth]{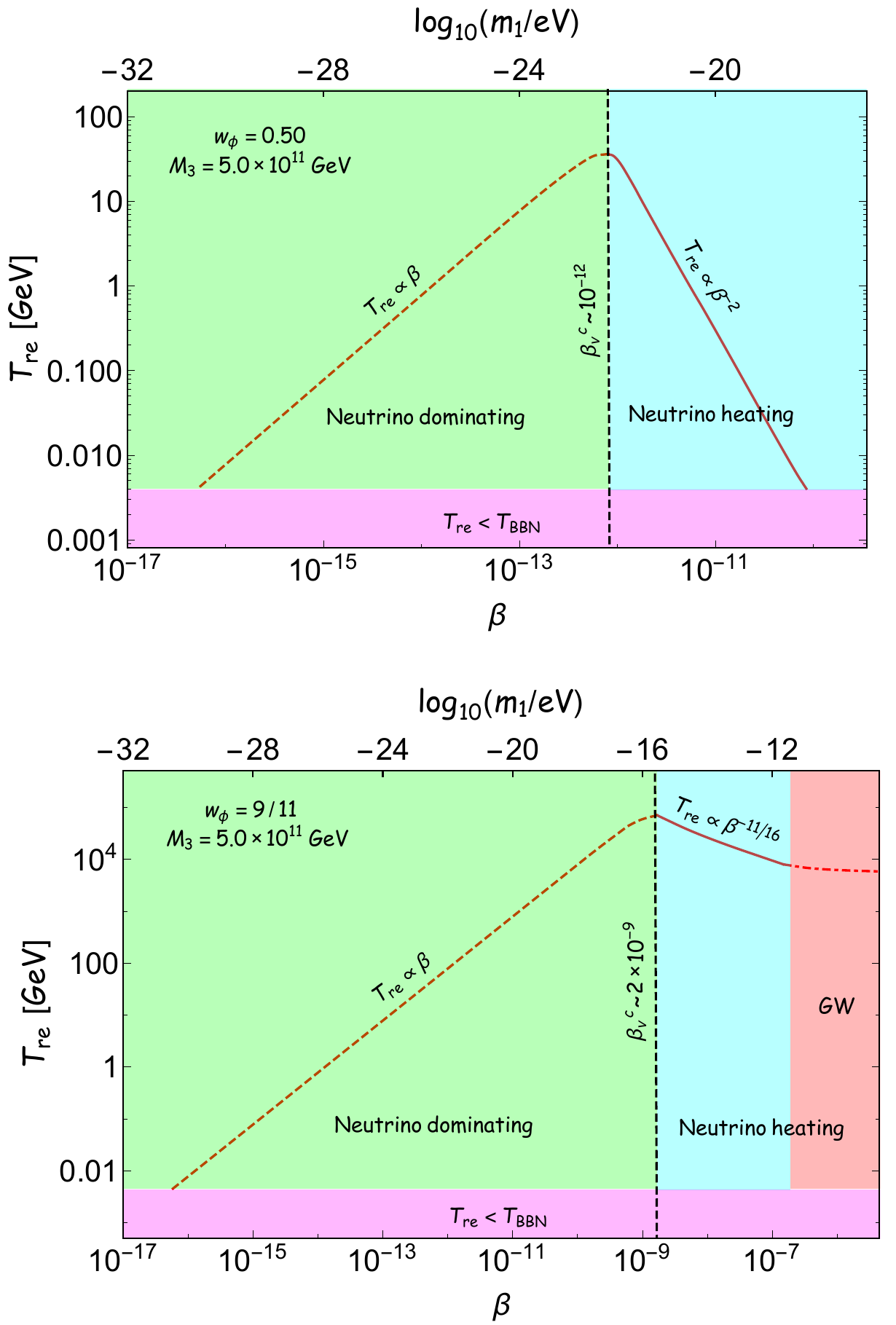}
\caption{Variation of reheating temperature $T_{\rm re}$ with $\beta$ for  $w_{\rm\phi}=1/2$ (top) and $w_{\rm\phi}=9/11$ (bottom). BBN bounds exclude the magenta-shaded regions, and red-shaded regions (bottom plot) are excluded by an excess of gravitational wave.}
\label{Tre}
\end{figure}
The final reheating temperature ($\tre$) is determined by $\nu^3_{\rm R}$ decay, as indeed can be seen in the top plot of Fig.\ref{density}, and this is realized only for $w_\phi\geq1/3$. The reheating ends at $\are$ where $\Gamma_{\rm 3}\sim H(a_{\rm re})$ is satisfied which leads to,

\begin{equation}{\label{TreN}}
\begin{aligned}
    T_{\rm re}
    &\simeq  10^{14}\beta\left(\frac{M_3}{5\times 10^{11}\mbox{GeV}}\right)^{1/2}\mbox{GeV}
    \end{aligned}
\end{equation}
Where, we utilized the relation $\rho_{\rm r} =\epsilon\, T^4$, for radiation temperature $T$ with $\epsilon=(\pi^2/30)g_{\rm \star r}$. The relativistic degrees of freedom in the radiation $g_{\rm \star r} \simeq 100$. For example if we assume $\tre = T_{\rm  BBN} = 4$ MeV, one gets $\beta \simeq 4\times10^{-17}$ for $  M_3=5\times10^{11}$ GeV. 
Further, the maximum possible $\beta=\beta^{\rm c}_\nu$ leads to maximum reheating temperature for a given $M_3$.
 If we assume $w_\phi = 9/11\,(1/2)$ and $M_3=5\times10^{11}$, the maximum temperature turns out to be $10^{5}\,(40)$ GeV.  
We indeed recover these values with  $T_{\rm re} \propto \beta$ relation numerically as shown in Fig.\ref{Tre} with the dashed line. 
 
\subsection{Neutrino heating case: \,$\beta^{c}_\nu\leq\beta\leq\beta^{c}_\phi$: }
For this case, whereas the inflaton controls the background reheating dynamics, the process of heating the thermal bath is still dominated by RHNs and determines the final reheating temperature. After the neutrino heating starts dominating, the radiation from $\nu_{\rm R}^3$ decay evolves as (solving Eq.\ref{rhor}
\begin{equation}\label{rhorphi}
    \rho_{\rm r}(a)\simeq\frac{2\beta^2_{\rm 3}M^2_{\rm 3}n^{\rm end}_{\rm3}}{8\pi(5+3w_{\rm\phi})H_{end}}\left(\frac{a}{a_{\rm end}}\right)^{-\frac{3(1-w_{\rm\phi})}{2}} .
\end{equation}
This evolution will continue till the instant say at $a=a_{\rm \nu H} \leq \are$, for which $\Gamma_{\rm 3}\sim H$ satisfied. 
 \begin{equation}
     \begin{aligned}
         \frac{a_{\rm \nu H}}{a_{\rm end}}=\left(\frac{\beta^2}{12\pi(1+w_{\rm\phi})}\frac{M_{\rm 3}}{H_{\rm end}}\right)^{\frac{-2}{3(1+w_{\rm\phi})}}.
     \end{aligned}
 \end{equation}
After this instant $a_{\rm \nu H}(<a_{\rm re})$, any additional entropy injection into the thermal bath will be ceased, and the radiation component falls as $a^{-4}$. With this fall off, reheating completes at $\rho_\phi(a_{\rm re})\sim \rho_{\rm r}(a_{\rm re})$ (see the bottom plot of Fig.\ref{density}), and it is possible only when the inflaton EoS $w_\phi>1/3$. {Hence, $\tre$ can be expressed as,
 \begin{equation}{\label{Trephi}}
\begin{aligned}
    &T_{\rm re}\simeq 0.5M_{\rm p}\,\beta^{\frac{-1}{3w_\phi-1}}\left(\frac{M_{\rm p}}{H_{\rm end}}\right)^{\frac{3}{2(3w_\phi-1)}}
    \left(\frac{n^{\rm end}_3}{M^3_{\rm p}}\right)^{\frac{3(1+w_\phi)}{4(3w_\phi-1)}}\\
&\quad\quad\quad\times\left(\frac{M_{\rm p}}{M_{\rm 3}}\right)^{\frac{1+3w_\phi}{4(1-3w_\phi)}}\,.
     \end{aligned}
 \end{equation}
 Reheating temperature depends nontrivially on the neutrino coupling, $\tre\propto\beta^{-1/{(3\wphi-1)}}$. For example, if we assume $w_\phi=1/2$, the $\tre\propto\beta^{-2}$, and for $w_\phi=9/11$, we got $\tre\propto \beta^{-11/16}$ which we recover numerically as shown in Fig.\ref{Tre} with solid lines. For this neutrino heating case, the maximum $\tre$ should be at $\beta = \beta^{\rm c}_\nu$, and it is the same as the neutrino dominating case where both meet in $(\beta, \tre)$-plane (see Fig.\ref{Tre}). The minimum $\tre$ is set by the GRe temperature if $\wphi>0.65$ (and $\beta =\beta^{\rm c}_{\phi}$), otherwise it would be $T_{\rm BBN}$.
 
 {\subsection{ Constraining inflation} Using same $M_3$ value, we have projected our results in $n_{\rm s}-r$ plane for two different $w_\phi=9/11,\,0.5$ (see top Fig.\ref{nsr}) and shown the prediction of maximum $\alpha$ value from the observational $68\%$ and
$95\%$ CL constraints from BICEP/Keck data. For $w_\phi=9/11,\alpha=1$, we found the  range of $n_s$ lying within $0.96491 \leq n_s \leq 0.96825$, which is well inside the $1\,\sigma$ bound. That in turn fixes inflationary e-folding number ($N_{\rm I}$) within $57\leq N_{\rm I}\leq 64$. Similar bounds can also be obtained for all allowed values of $\alpha$. 
\begin{figure}[t]
\centering
\includegraphics[width=\columnwidth]{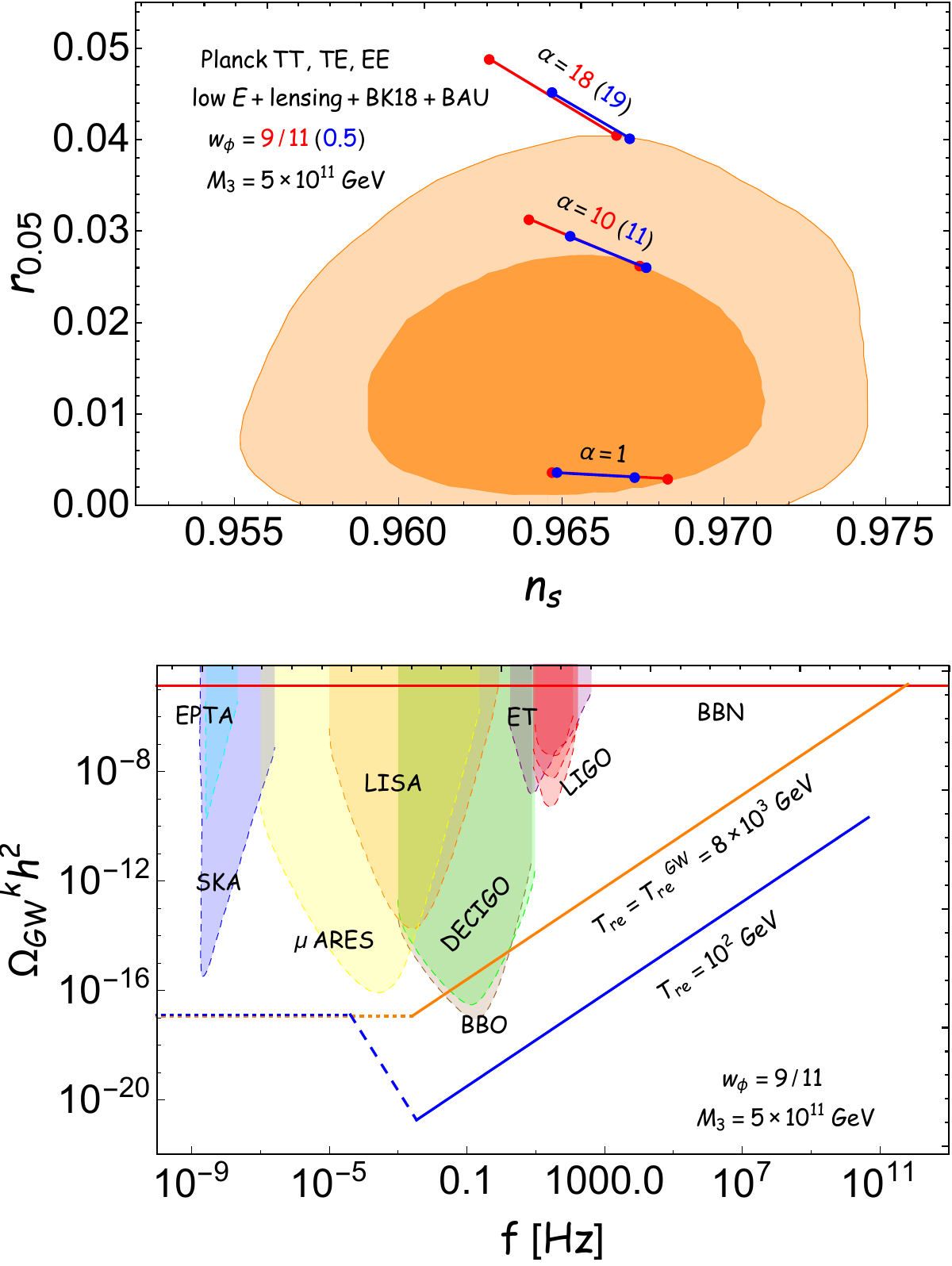}
\caption{\textbf{Top}: Compare our
result with the observational $68\%$ and $95\%$  CL constraints from BICEP/Keck, in the $(n_{\rm s}, r)$ plane. \textbf{Bottom}: Behavior of $\Omega_{\rm GW}^{\rm k}h^2$ over a wide range of frequency $f=k/2\pi.$ 
}
\label{nsr}
\end{figure}

\subsection{PGW Constraints}
One of the profound predictions of inflation is the existence of PGWs \cite{Grishchuk:1974ny,Guzzetti:2016mkm,Starobinsky:1979ty}. It acts as a unique probe of the early universe.
The amplitude and the evolution of the PGWs spectrum are sensitive to the energy scale of the inflation and the post-inflationary expansion. We are particularly interested in those modes between $k_{\rm re} < k < k_{\rm end}$, which re-enter the horizon during reheating. Depending on the post-inflationary expansion, the PGW spectrum traces out the features of those phases.
The PGW spectral tilt generically assumes the form $ n_{\rm GW} =\frac{6w-2}{1+3w}$ \cite {Mishra:2021wkm,Haque:2021dha} which predicts a blue-tilted spectrum for an EoS $w>1/3$ and red-tilted for $w<1/3$.

\underline{Neutrino dominating case}:
For this case, there is an intermediate $\nu^3_{\rm R}$ dominated matter-like phase ($w=0$).  Corresponding to this phase there exists a particular scale $k_\nu =a_\nu H(a_\nu)$, that re-enters the horizon at the inflaton and neutrino equality point $a_\nu$. Where $k_{\rm end}=e^{N_{\rm I}}k_{\rm *}$ is the mode that enters the horizon at the beginning of reheating and is related to the inflationary e-folding number $N_{\rm I}$ calculated at the CMB pivot scale $k_\star=0.05~ \mbox{Mpc}^{-1}$. 
Since $\nu^3_{\rm R}$ domination behaves as dust,  following the expression for the spectral tilt mentioned before, the present-day PGWs spectrum corresponding to those modes lies in between $k_{\rm \nu}>k>k_{\rm re}$ will be, 
\be \label{GW2}
\Omega^{\rm k}_{\rm GW}h^2\simeq 
3.4 \times 10^{-18} \left(\frac{k}{ k_{\rm re}}\right)^{-2}\,\left(\frac{H_{\rm end}}{10^{13}\,\mbox{GeV}}\right)^2 .
\ee
Where $k_{\rm re}=a_{\rm re}H(a_{\rm re})$ is the mode that enters the horizon at the end of reheating, and its value naturally depends on $T_{\rm re}$ as,
\begin{equation}
\begin{aligned}
   \kre
   &\simeq 1.2\times10^{-9}\,\left(\frac{\tre}{4~ \mbox{MeV}}\right)\, \mbox{Hz} .
   \end{aligned}
\end{equation}
Due to the red tilted spectrum with $k^{-2}$, as shown in the blue dashed line in the bottom plot of Fig.\ref{nsr}.
On the other hand, modes within  $\kend \geq k \geq k_\nu $ enter the horizon during the early inflaton-dominated phase, and the PGWs spectrum assumes the following form,
\be \label{GW3}
\Omega^{\rm k}_{\rm GW}h^2\simeq 1.9\times 10^{-18}\left(\frac{H_{\rm end}}{10^{13}\,\mbox{GeV}}\right)^2 
\left(\frac{k_{\rm re}}{k_\nu}\right)^{2} \left(\frac{k}{ k_{\rm \nu}}\right)^{\frac{(6\,w_\phi-1)}{(1+3\,w_\phi)}}  .
\ee
PGW spectrum is always shown to be blue tilted as reheating happens for $w_{\rm\phi}\geq1/3$. For $w_\phi=9/11$, the $\Omega^{\rm k}_{\rm GW}h^2$ behaves as $k^{16/19}$ (see solid blue line in bottom plot of Fig. \ref{nsr}).

\underline{Neutrino heating case:}
For this case, inflaton is the dominant component during the whole reheating period. All the modes that enter the horizon during reheating, the PGWs spectrum at the present time can be written as \cite{Haque:2021dha,Chakraborty:2023ocr} 
\be \label{GW}
\Omega^{\rm k}_{\rm GW}h^2\simeq 
1.9 \times 10^{-18} \left(\frac{k}{ k_{\rm re}}\right)^{-\frac{(2-6\,w_\phi)}{(1+3\,w_\phi)}}\,\left(\frac{H_{\rm end}}{10^{13}\,\mbox{GeV}}\right)^2 .
\ee
Here, the PGW spectrum is also shown to be blue tilted (solid orange line). The PGWs spectrum would be maximum when $k=k_{\rm end}$, and it is bounded by the BBN bound of $\Omega^{\rm k_{end}}_{\rm GW}h^2\leq1.7\times10^{-6}$ \cite{Pagano:2015hma,Yeh:2022heq,Planck:2018vyg} which is depicted by red straight line. 
Therefore, using the BBN bound of PGW strength 
one obtains a
lower bound on the reheating temperature for $w_\phi>1/3$ as,
\begin{equation}{\label{tregw}}
    \tre>0.5M_{\rm p}(2.4\,\pi)^{\frac{3(1+w_\phi)}{4(1-3w_\phi)}}\left(\frac{H_{\rm end}}{M_{\rm p}}\right)^{\frac{1+3w_\phi}{3w_\phi-1}} .
\end{equation}
Setting the above temperature with the BBN energy scale
$\sim4\,\mbox{MeV}$, we can see that the BBN bound of
PGWs are only important when $w_\phi\geq0.60$. Moreover, note that to satisfy this restriction on the reheating temperature, we can always estimate a $\beta$ value (see, for instance, Eq.\ref{Trephi}) below which our analysis is true.   

\section{ Leptogenesis and constraints}{\label{sc6}}
Since RHNs are gravitationally produced from the inflaton, they will undergo CP-violating out-of-equilibrium decay and produce lepton asymmetry. By the well-known non-perturbative sphaleron processes \cite{sy,ja}, those lepton asymmetries are then converted into baryon asymmetry. For our analysis we  considered the following mass hierarchy $M_{1}\lesssim m^{\rm end}_{\rm\phi}\ll M_2$. 
The  CP asymmetry parameter $(\epsilon_{\Delta L})$ generated from the decay of $\nu^1_{\rm R}$, using the seesaw mechanism  \cite{Buchmuller:2004nz,Barman:2022qgt,Barman:2023opy,Co:2022bgh} is expressed as , 
\begin{equation}
\begin{aligned}
\epsilon_{\Delta L}
\simeq \frac{3\delta_{\rm eff}}{16\pi}\frac{m_3\,M_{1} }{v^2}
\simeq9.85\times10^{-5}\left(\frac{M_1}{10^{12}\,\mbox{GeV}}\right)
\end{aligned}
\end{equation}
The
effective CP-violating phase ($\delta_{\rm eff}$) in the neutrino mass matrix is assumed 
to be $\delta_{\rm eff}=1$ for our present analysis. 
\begin{equation}{\label{yb}}
    Y_{\rm B}=\frac{n_{\rm B}}{s}=\frac{28}{79}\epsilon_{\Delta L}\frac{n_{1}(T_{\rm re})}{s(T_{\rm re})}\, \simeq 8.7\times 10^{-11},
\end{equation}
where $s(T_{\rm re})=\frac{2\pi^2}{45}g_{\rm\star s}T^3_{\rm re}=\frac{4\,\epsilon}{3}\frac{g_{\rm\star s}}{g_{\rm\star r}}T^3_{\rm re}$ is the entropy density at the end of reheating. Where $g_{\rm \star s}$ counts the number of entropic degrees of freedom. The number density, $n_{1}(T_{\rm re})$ at the time of reheating from Eq.\ref{Nis}, 
\begin{equation}\label{Nisre}
    n_{\rm 1}(\tre)= \frac{3H^3_{\rm end}M^2_{\rm 1}}{2\pi(1-w_\phi)(10\,\gamma \,m^{\rm end}_\phi)^2} \left(\frac{\are}{\aend}\right)^{-3}\,.
\end{equation}
We will now use this expression
in Eq. \ref{yb}, to constrain our model parameters. Utilizing the expression of RHN number density for two different reheating scenarios, we arrived at the final exclusion plots in $(w_\phi, \beta)$ plane (see Fig.\ref{betavswphi}).   
\begin{figure}[t]
\centering
\includegraphics[width=\columnwidth]{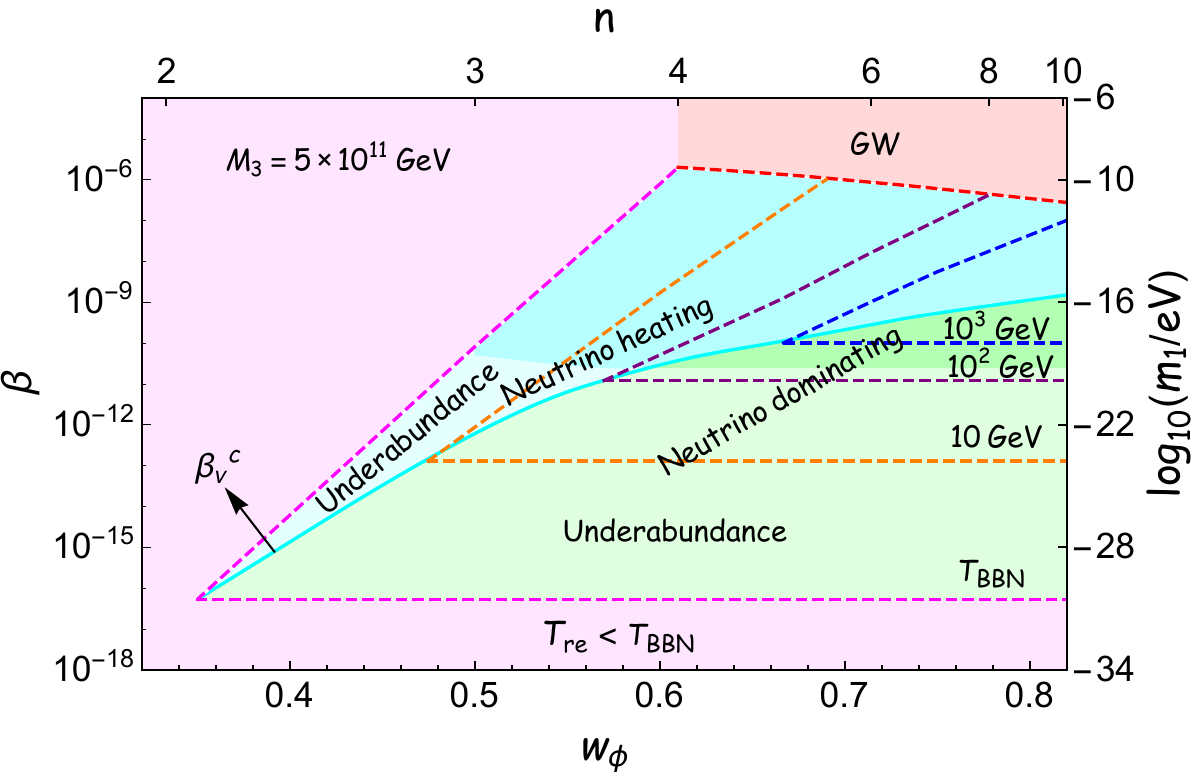}
 \caption{$\beta$ versus $w_{\rm\phi}$ parameter space for a given $M_3=5\times10^{11}$ GeV.  The deep cyan-shaded and green-shaded regions correspond to the regions where the observed baryon asymmetry is possible. The light cyan-shaded and green-shaded regions lead to the underabundance of $Y_{\rm B}$. }
\label{betavswphi}
\end{figure}

The baryon asymmetry provides the strongest constraint on our reheating scenario. The green-shaded region in Fig.\ref{betavswphi} corresponds to the neutrino dominating case, and the cyan-shaded region corresponds to the neutrino heating case. Particularly, the deep green and deep cyan shaded regions are where right baryon asymmetry is produced. On the other hand, the light green and light cyan-shaded regions represent the under-abundant baryon asymmetry. The magenta and red-shaded regions are ruled out from the BBN bound of the reheating temperature and PGWs, respectively. 

For Fig.\ref{betavswphi}, we particularly considered $M_3=5\times 10^{11}$ GeV. Combining two reheating cases, we found the overall allowed equation of state should be $w_\phi \gtrsim 0.5$, for which the right baryon asymmetry is generated, and the lightest active neutrino mass is predicted to be within $(10^{-10} \gtrsim m_1 \gtrsim 10^{-19})$ eV. Although we have shown all the results for a particular $\nu^3_{\rm R}$ mass, rather wider mass ranges of $\nu^3_{\rm R}$ within $10^{10} -  10^{14}~\mbox{GeV}$ is observed to be allowed. And corresponding  mass range of $\nu^1_{\rm R}$ for successful leptogenesis is within $M_1\simeq(8-350)\times10^{11}$ GeV.
\section{Conclusions}{\label{scl}}

In summary, we proposed a new reheating mechanism by invoking the Type-I seesaw neutrino model, which simultaneously resolves the SM neutrino mass and baryon asymmetry problems and also offers a successful reheating scenario. This reheating turned out to further constrain the inflation models within a very narrow range in the $n_{\rm s}-r$ plane. We found two possible scenarios. If the reheating is controlled by neutrino domination, we found the reheating temperature $T_{\rm re}\propto\beta \, M_3^{1/2}$, and reaches its maximum at $\beta=\beta_{\nu}^c$. For example, if $w_\phi = 9/11$, the maximum temperature turned out to be as high as $10^{6}$ GeV,
and $\beta$ shoud be $> 10^{-17}$ with fixed by $T_{\rm BBN}$. For this case within $\kre < k < k_\nu$, GW spectrum behaves as $\Omega^{\rm k}_{\rm GW}h^2 \propto k^{-2}$, and for $\kend < k < k_\nu$, the spectrum behaves as $\Omega^{\rm k}_{\rm GW}h^2 \propto k^{-\frac{(2-6\,w_\phi)}{(1+3\,w_\phi)}}$ which is as expected for the inflaton dominated background. \\
On the other hand, for neutrino heating scenario, relation $T_{\rm re} \propto \beta^{\frac{-1}{3w_\phi-1}}
  M_{\rm 3}^{\frac{7+9w_\phi}{4(3w_\phi-1)}}$ leads to $\tre$ decreasing with $\beta$ unlike the previous case. 
  Since inflaton is the dominating component during reheating, the PGW spectrum behaves as $\Omega^{\rm k}_{\rm GW}h^2 \propto k^{-\frac{(2-6\,w_\phi)}{(1+3\,w_\phi)}}$.

\section{acknowledgements}
RM would like to thank the Ministry of Human Resource Development, Government of India (GoI),
for financial assistance. MRH wishes to acknowledge support from the Science and Engineering Research Board (SERB), Government
of India (GoI), for the SERB National Post-Doctoral fellowship, File Number: PDF/2022/002988. DM wishes
to acknowledge support from the Science and Engineering Research Board (SERB), Department of Science and
Technology (DST), Government of India (GoI), through the Core Research Grant CRG/2020/003664. RM would like to thank Niloy Mondal for the helpful discussions. DM and RM also thank the Gravity and High Energy Physics groups at IIT Guwahati for illuminating discussions.

\hspace{0.5cm}

\end{document}